\documentclass{article}

\usepackage{arxiv}

\usepackage[utf8]{inputenc} 

\usepackage{amsfonts}       
\usepackage{nicefrac}       
\usepackage{microtype}      
\usepackage{lipsum}

\usepackage{color}
\usepackage{listings}
\usepackage{booktabs}
\usepackage{array}
\usepackage{graphicx}
\usepackage{longtable}
\usepackage{subfigure}

\usepackage{cite}
\usepackage{url}
\usepackage{fancyhdr}

\usepackage{soul}

\usepackage{float}
\usepackage{listings}
\usepackage{mdwmath}
\usepackage{mdwtab}
\usepackage{multirow}
\usepackage{multicol}
\usepackage{rotating}
\usepackage{setspace}
\usepackage[utf8]{inputenc}
\usepackage{lineno}
\usepackage{listings}

\usepackage{mdwmath}
\usepackage{mdwtab}
\usepackage{multirow}
\usepackage{multicol}
\usepackage{array}
\usepackage{booktabs}

\usepackage{enumitem}
\usepackage{xspace}
\usepackage[export]{adjustbox}
\usepackage{graphicx}
\usepackage{color,soul}
\usepackage{rotating}
\usepackage{setspace}
\usepackage{amsmath} 
\usepackage{amssymb}
\usepackage{float}
\usepackage{xcolor}

\usepackage{hyperref}
\usepackage[numbers]{natbib}
\usepackage{url}

\title{Developing Responsible Chatbots for Financial Services: A Pattern-Oriented Responsible AI Engineering Approach}

\author{Qinghua Lu\textsuperscript{1}, 
Yuxiu Luo\textsuperscript{2},
Liming Zhu\textsuperscript{1},
Mingjian Tang\textsuperscript{3},
Xiwei Xu\textsuperscript{1},
Jon Whittle\textsuperscript{1}\\
\textsuperscript{1}Data61, CSIRO, Australia\\
\textsuperscript{2}Westpac, Australia\\
\textsuperscript{3}Atlassian, Australia
}

\begin{document}

\maketitle

\begin{abstract}
The recent release of ChatGPT has gained huge attention and discussion worldwide, with responsible AI being a key topic of discussion. How can we ensure that AI systems, including ChatGPT, are developed and adopted in a responsible way? To tackle the responsible AI challenges, various ethical principles have been released by governments, organisations, and companies. However, those principles are very abstract and not practical enough. Further, significant efforts have been put on algorithm-level solutions that only address a narrow set of principles, such as fairness and privacy. To fill the gap, we adopt a pattern-oriented responsible AI engineering approach and build a Responsible AI Pattern Catalogue to operationalise responsible AI from a system perspective. In this article, we first summarise the major challenges in operationalising responsible AI at scale and introduce how we use the Responsible AI Pattern Catalogue to address those challenges. We then examine the risks at each stage of the chatbot development process and recommend pattern-driven mitigations to evaluate the the usefulness of the Responsible AI Pattern Catalogue in a real-world setting.

\end{abstract}

\section{Introduction}
ChatGPT has gained huge attention and discussion worldwide, with responsible AI being a crucial topic of discussion. One key question is how we can ensure that AI systems, like ChatGPT, are developed and adopted in a responsible way? 
Responsible AI is the practice of developing, deploying, and maintaining AI systems in a way that benefits the humans, society, and environment, while minimising the risk of negative consequences. 
To solve the challenge of responsible AI, many AI ethics principles have been released recently by governments, organisations, and enterprises~\cite{jobin2019global}. 

A principle-based approach provides technology-neutral and context-independent guidance while allowing context-specific interpretations for implementing responsible AI. However, those principles are too abstract and high-level for practitioners to use in practice. For example, it is a very challenging and complex task to operationalise the the human-centered value principle regarding how it can be designed for, implemented and monitored throughout the entire lifecycle of AI systems. In addition, the existing work mainly focuses on algorithm-level solutions for a subset of mathematics-amenable AI ethics principles (such as privacy and fairness). However, responsible AI issues can happen at any stage of the development lifecycle crosscutting various AI and non-AI components of systems beyond AI algorithms and models. To try to bridge the principle-algorithmic gap, further guidance such as guidebooks\footnote{\url{https://www.microsoft.com/en-us/haxtoolkit/}}\footnote{\url{https://pair.withgoogle.com/guidebook}}, question banks,~\cite{liao2020questioning}, checklists~\cite{han2022checklist} and documentation templates~\cite{raji2020closing,hutchinson2021towards} have begun to emerge. Those attempts tend to be ad-hoc and lack of systematic solutions to cover the entire lifecycle of AI systems taking into account different levels of stakeholders. 

We have adopted a pattern-oriented responsible AI engineering approach~\cite{lu2022towards} and built a Responsible AI Pattern Catalogue \footnote{\url{https://research.csiro.au/ss/science/projects/responsible-ai-pattern-catalogue/}} for different types and levels of stakeholders in AI industry~\cite{lu2022responsible, lu2023responsible}. In this article, we first summarise the major challenges in operationalising responsible AI at scale and introduce how responsible AI pattern catalogue addresses those challenges. Then, we examine the risks at each stage of the chatbot development process and recommend pattern-oriented mitigations to evaluate the usefulness of the Responsible AI Pattern Catalogue. 

\section{RELATED WORK}

The concept of chatbots can be traced back to the 1950s when computer scientist and inventor Alan Turing proposed the Turing Test, which aimed to determine whether a machine could exhibit intelligent behavior indistinguishable from a human.
In 1966, Joseph Weizenbaum created ELIZA, the first known chatbot, which was designed to simulate a psychotherapist by responding to user inputs with pre-programmed responses~\cite{weizenbaum1976computer}.
In the 1980s and 1990s, advances in natural language processing and machine learning led to the development of more advanced chatbots, such as Parry and ALICE.
In the early 2000s, the rise of messaging platforms and mobile devices made it easier for businesses to integrate chatbots into their customer service systems.
In recent years, advancements in AI, such as deep learning and natural language processing, have made it possible for chatbots (e.g., OpenAI's ChatGPT\footnote{\url{https://openai.com/blog/chatgpt/}}, IBM Watson Assistant\footnote{\url{https://cloud.ibm.com/catalog/services/watson-assistant}}) to handle more complex and natural conversations with users, leading to the widespread use of chatbots in various industries, including finance, healthcare, and e-commerce. 
Despite the increasing popularity of chatbots, people have many ethical concerns about chatbots. Some studies on chatbots have begun to emerge, such as human trust and emotion~\cite{shin2022cross, shin2022perception, shin2022effects}. Significant efforts have been put on algorithm-level solutions which mainly focus on a subset of ethical principles~\cite{shin2022algorithm, shin2022understanding, shin2022platforms, shin2022seeing}, such as privacy, fairness, and explainability. However, there is lack of responsible AI governance and engineering studies to assess and mitigate the ethical risks of chatbots against all the AI ethics principles.


\section{MAJOR CHALLENGES IN OPERATIONALISING RESPONSIBLE AI AT SCALE}

Through our engagement with industry, we have summarised three major challenges in operationalising responsible AI at scale: 
\begin{itemize}
    \item \textbf{Challenge 1: Diverse Stakeholders and Risk Landscape.} Organisations usually take a risk-based approach to ensuring responsible AI. However, in AI ecosystems, different levels of stakeholders have various interests on responsible AI risks. Industry-level stakeholders (e.g., responsible AI regulators) and organisation-level stakeholders (e.g., board and executives) are more interested in harms and societal impacts, associated preventive/corrective cost and governance mechanisms, while team-level stakeholders (e.g., developers) care more about algorithmic risks, reliability and security techniques. AI technology/solution procurers are a type of industry-level stakeholders who are interested in how to verify the ethical quality of third party AI technologies/solutions without having full access to their proprietary models  and how to deal with accountability and contestability when third parties are involved. Those varying stakeholder interests in responsible AI risk management demanding different but connected methods and tools.  
    \item \textbf{Challenge 2: Competing Risk Silos.} Organisations usually create individual board risk committees to manage different types of risks, such as financial, reputation and legal risks with some dedicated cybersecurity and privacy risk management. However, having dedicated risk committees including responsible AI risk committees give rise to risk silos which usually have limited connections between risks assessed separately and compete for resources. This makes risk management perceived as a forced and meaningless roll-up in organisations.
    \item \textbf{Challenge 3: Lack of Risk Expertise.} Each organisation has existing and different governance and risk approaches. There is a shortage of expertise to assess new risks, including responsible AI risks. Most of the organisations do not have the necessary scope to handle responsible AI risks and the capacity to examine each AI project deeply. Thus, the organisations heavily reply on the project teams to do self-assessment. The current risk assessment practices include checklists, conversations, information sheets, rather than formal or technical approaches. Organisations usually treat risk analysis as hazard/threat analysis, omitting system vulnerability, exposure risks, and response/mitigation risk.
    
\end{itemize}

\begin{figure*}
\centering
\includegraphics[width=0.75\textwidth]{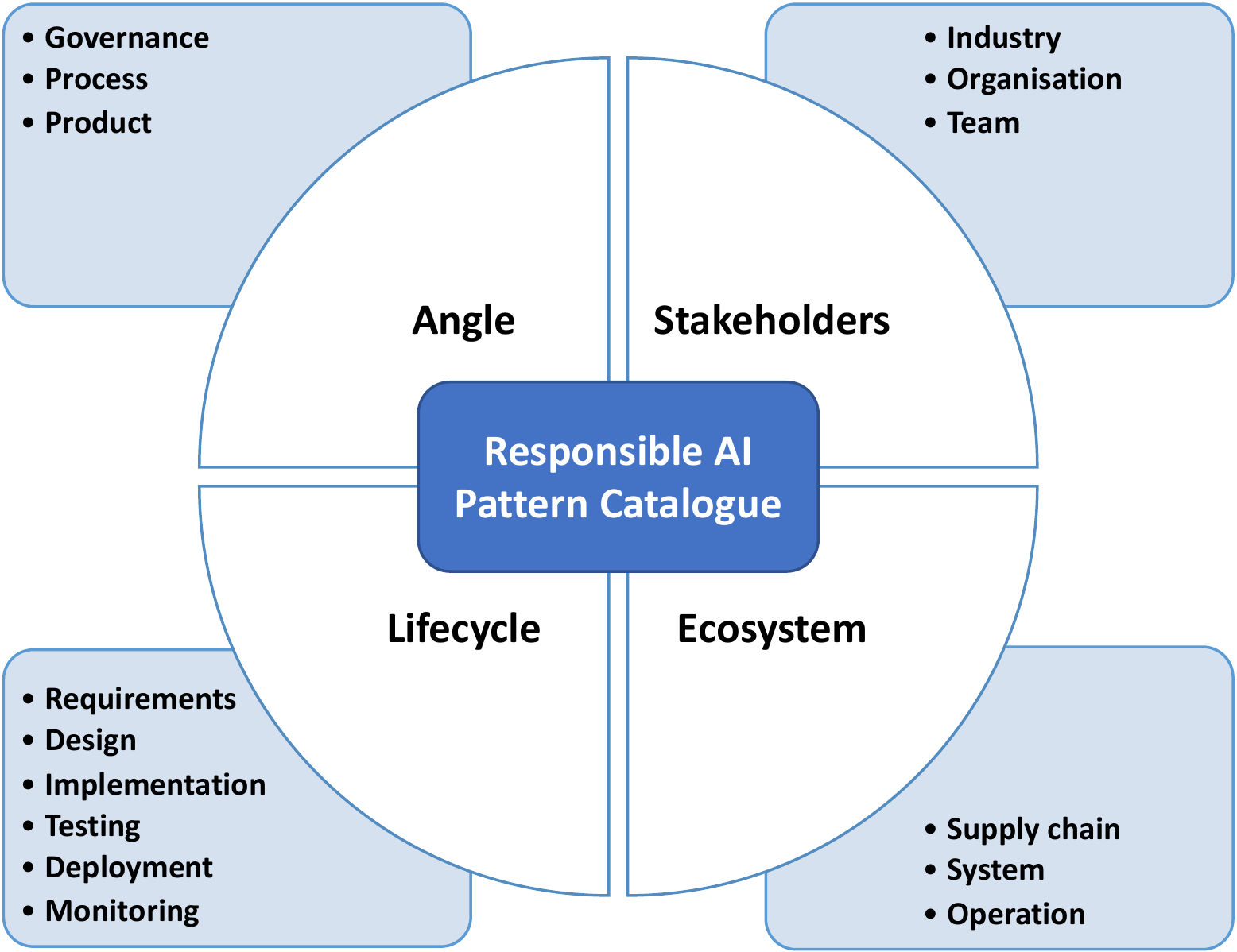}
\caption{Overview of responsible AI pattern catalogue.} \label{fig:catalogue}
\vspace{-2ex}
\end{figure*}

\begin{figure*}
\centering
\includegraphics[width=\textwidth]{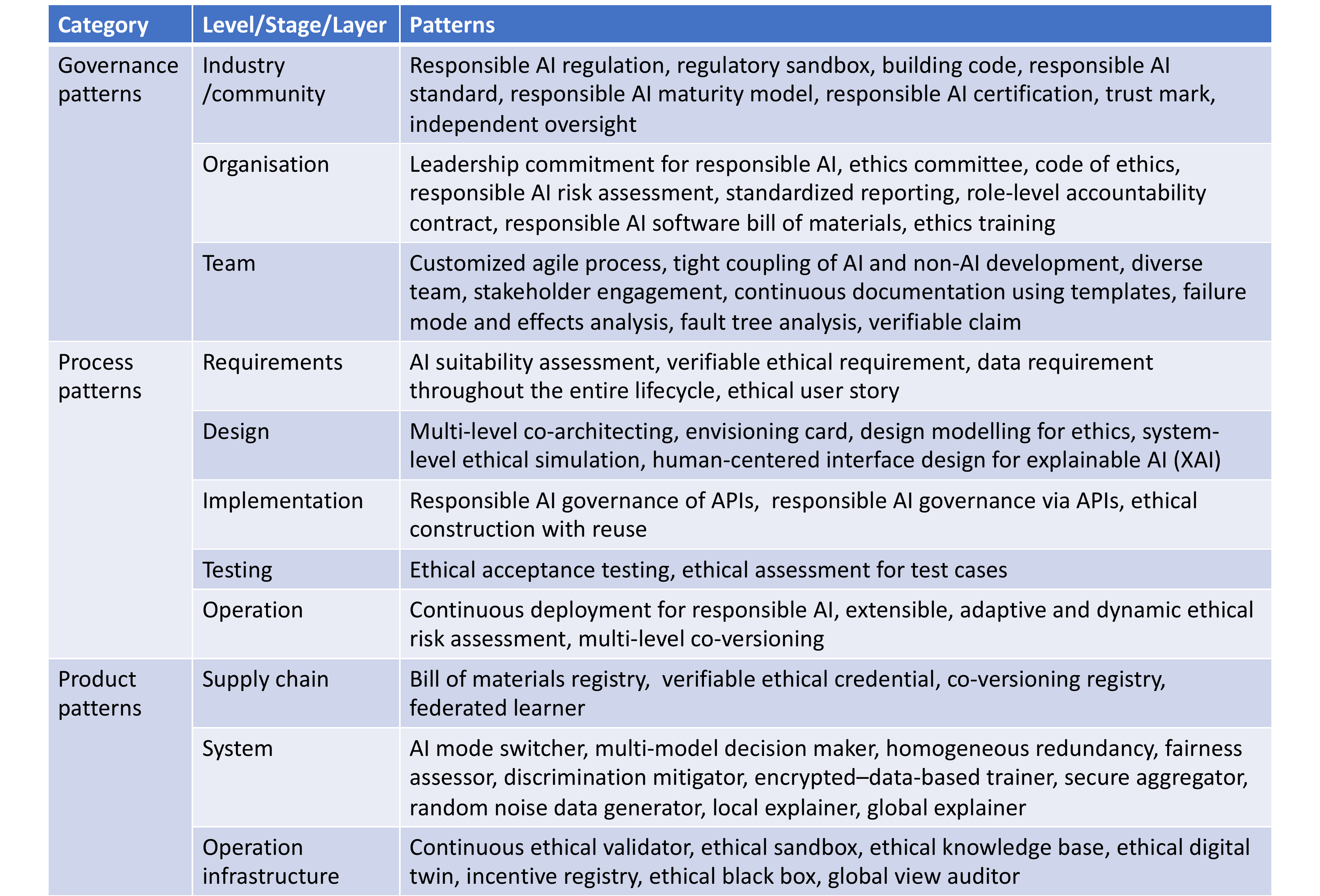}
\caption{List of responsible AI patterns.} \label{fig:patterns}
\vspace{-2ex}
\end{figure*}

\section{PATTERN-ORIENTED RESPONSIBLE AI ENGINEERING APPROACH: RESPONSIBLE AI PATTERN CATALOGUE}

We have adopted a pattern-oriented responsible engineering approach and built a responsible AI pattern catalogue to address the end-to-end/top-to-bottom responsible AI challenges~\cite{lu2022responsible, lu2023responsible}. 
In software engineering, a pattern is a reusable solution to a recurring problem within a given context during software development. Patterns are documented to capture the knowledge about reusable solutions in an accessible and structure way for stakeholders to learn. A pattern catalogue is a collection of patterns that are related to some extend and can be used together or independently.

Based on the results of a multivocal review~\cite{lu2022responsible}, we analysed successful case studies and generalised best practices into patterns. The current version of pattern catalogue has collected 63 patterns, including 24 governance patterns, 17 process patterns, and 22 product patterns. To describe the pattern, we extended the traditional pattern template with additional elements, including summary, pattern type, objective, target users, impacted stakeholders, lifecycle stages relevant AI ethics principles, context, problem, solution, consequences (i.e., benefits and drawbacks), related patterns and known uses. Each pattern provides meaningful analysis on consequences, where pointers are added to measurement metrics and methods, residues of risk, and new risk introduced. Patterns are connected at multi-levels, multi-angles and across AI system life cycles through the related patterns. The defined fields in the pattern template can help efficiently selecting the patterns for mitigating a certain risk.

As illustrated in Fig.~\ref{fig:catalogue}, the responsible AI pattern catalogue has the following characteristics to help stakeholders better navigate the landscape and achieve responsible AI systems more successfully. 
\begin{itemize}
    \item \textbf{Across multiple angles and connected – governance, process, and product.} As listed in Fig.~\ref{fig:patterns}, the patterns are organised into three interconnected categories for easier adoption for impact. 
    \begin{itemize}
    \item Governance patterns for building multi-level governance for responsible AI; 
    \item Process patterns for establishing responsible software development processes and AI engineering;
    \item Product patterns for building responsible-AI-by-design into AI systems~\cite{lu2022responsible1}.
    \end{itemize}
    The stakeholders should use product patterns as product features to enforce responsible AI principles directly in the product and verify/validate the product. In the meantime, the stakeholders should also use process and governance patterns to complement responsible AI further.
    \item \textbf{Across multiple organization levels and connected – industry/community, organisation, and teams.} The patterns introduced are at different levels, so stakeholders can situate the practice areas in the bigger picture and see how the patterns can fit in and how different patterns influence and reinforce each other from an industry/community, organisation, and team level.
    \item \textbf{Across system lifecycle and connected – requirements, design, implementation, testing, deployment, and post-deployment monitoring.} Across the lifecycle of AI systems, different process patterns can be applied at other times, with the outputs of one pattern becoming the input of another.
    \item \textbf{Across the supply chain, system, and operation layer and connected.} We connect the product patterns through a system reference architecture across AI supply chain, AI system, and operation/deployment infrastructure layer. 
    \item \textbf{Benefiting multiple connected risks.} Individual responsible AI risks should be managed in silos by using risk-specific solutions. The patterns often help multiple risks together to raise the responsible AI posture of the organization significantly. 
    \item \textbf{Acknowledging drawbacks and additional risks introduced.} Adopting pattern-oriented risk mitigation may introduce additional risks and costs. We recognize them by incorporating drawbacks in the patterns and connecting with other related ways to further address the challenges.
    \item \textbf{Clear differentiation of trust and trustworthiness.} Trustworthiness is the ability of an AI system to meet responsible AI principles, while trust is stakeholders' subjective estimates of the trustworthiness of the AI system. We recognize that the importance of gaining stakeholder trust goes beyond the objective trustworthiness of the systems\cite{zhu2022ai}. Gaining trust is about diverse and inclusive engagement, setting realistic expectations, and communicating trustworthiness evidence in a way that stakeholders can understand and meaningfully critique. We include trust and trustworthiness as pattern objectives.
\end{itemize}

\begin{figure*}
\centering
\includegraphics[width=0.85\textwidth]{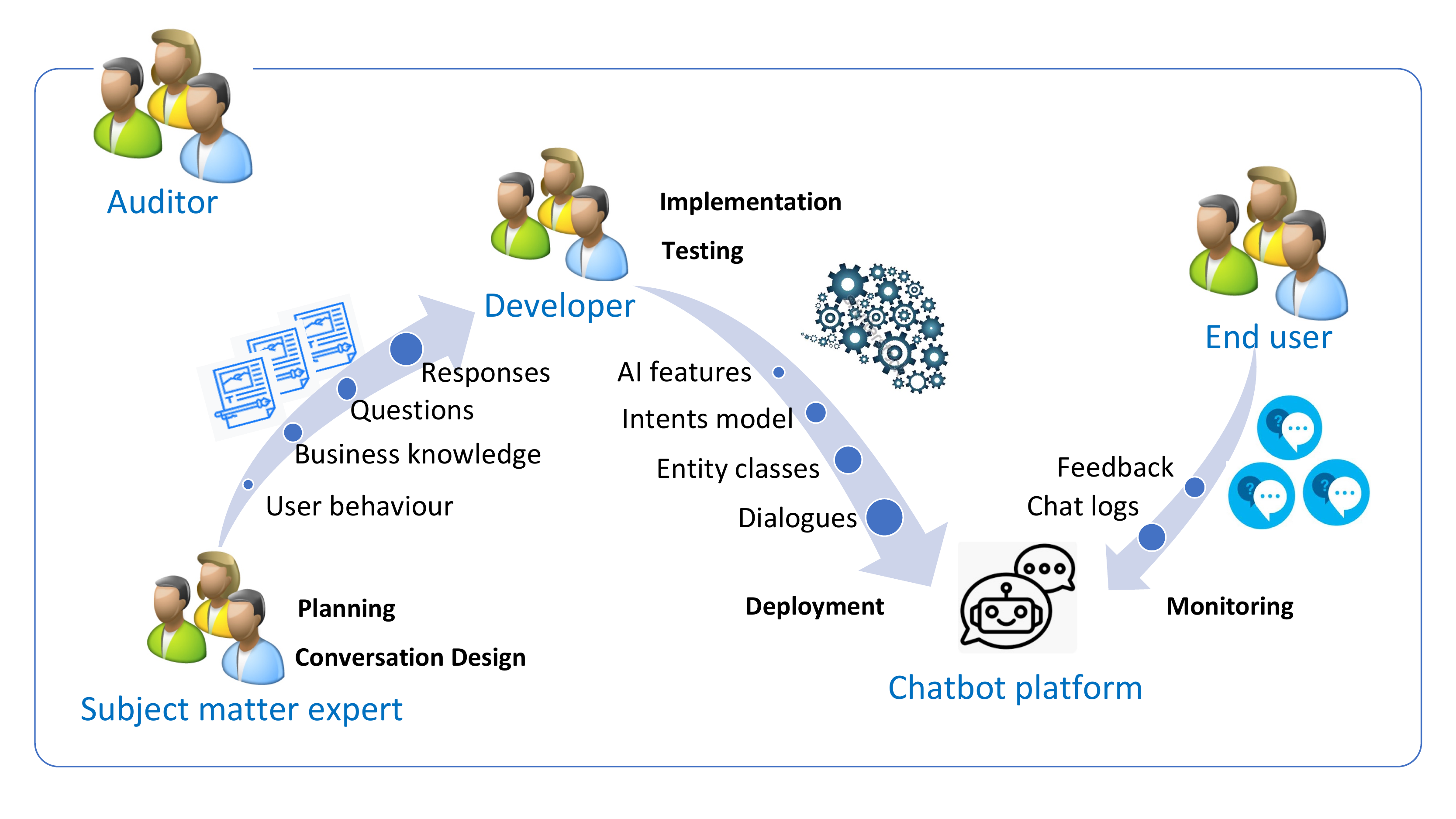}
\caption{Stakeholders and stages in the chatbot development.} \label{fig:stakeholders}
\vspace{-2ex}
\end{figure*}

\section{CASE STUDY: DEVELOPING CHATBOTS FOR THE FINANCIAL SERVICES}

\begin{figure*}
\centering
\includegraphics[width=\textwidth]{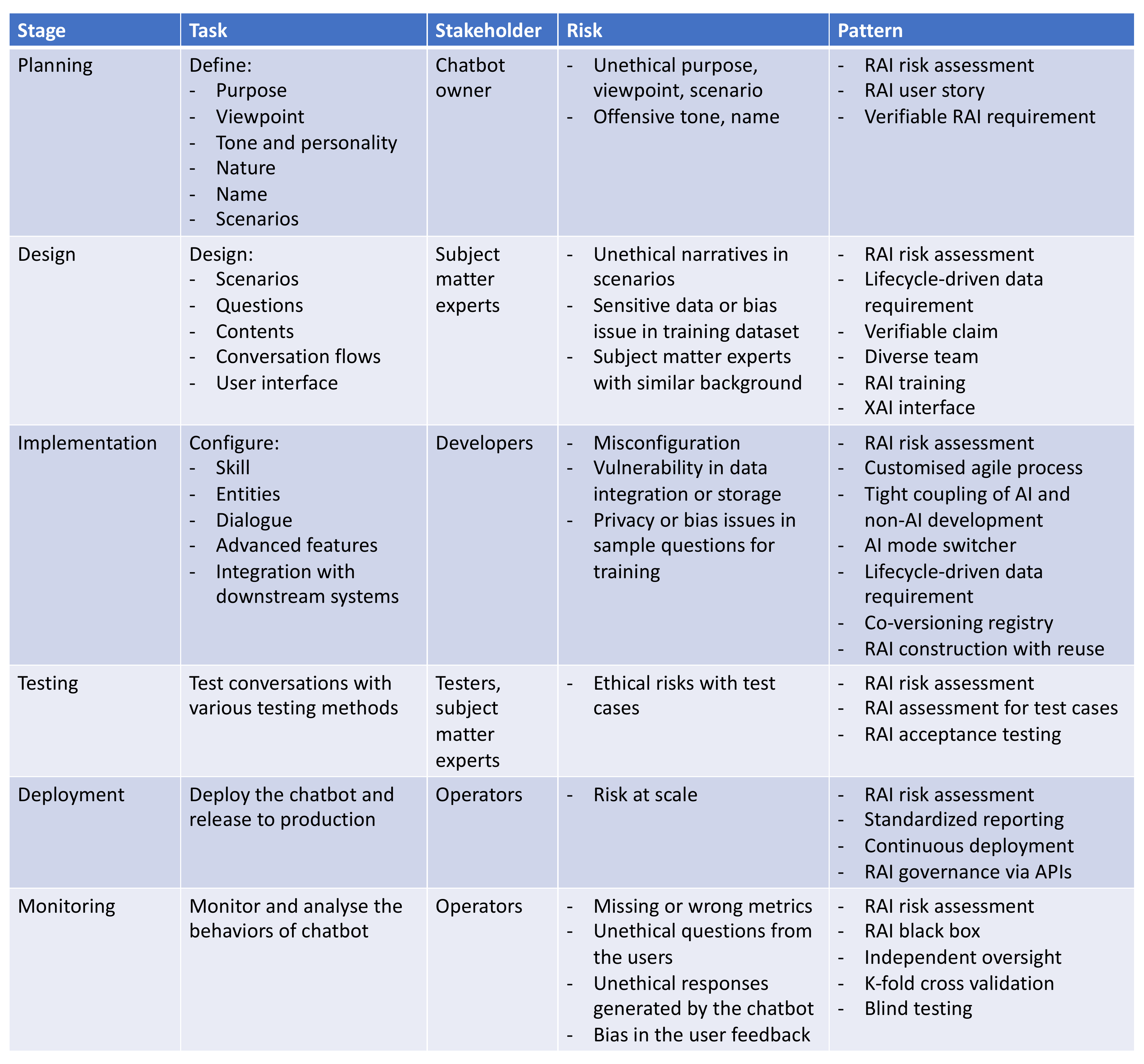}
\caption{Patterns for addressing risks in chatbot development.} \label{fig:risk}
\vspace{-2ex}
\end{figure*}

\subsection{Development Process of Chatbots}
In this section, we introduce the development process of chatbots using IBM Watson Assistant and discuss how patterns can be used to address various responsible AI risks, as illustrated in Fig.~\ref{fig:risk}. In the Responsible AI Pattern Catalogue, each pattern includes the specification of target users, lifecycle stages, and context, which can be used to locate the relevant patterns.

The chatbot development starts with the planning stage. Once the planning is completed, a conversation is designed and then built using the chosen chatbot platform, such as IBM Watson Assistant. The chatbot is then tested and deployed into target production environments. The performance of the chatbot is continuously measured. The chatbot is improved based the performance assessment and new requirements from business.  

\textit{Responsible AI risk committee pattern} can be adopted by the organisation to continuously assess the responsible AI risks of AI projects including chatbot projects through a dedicated responsible AI risk committee or the existing risk committee.
\textit{Responsible AI risk assessment pattern} can be used by the risk committee and project team to continuously assess the potential risks that could happen at each stage of the lifecycle of chatbots.

\subsubsection{Planning}

At the planning stage, stakeholders such as business representatives and chatbot owners need to discuss the following aspects: 
\begin{itemize}
    \item Overall purpose, such as customer service and internal help desk.
    \item Viewpoint which supports the purpose of the chatbot and provides a consistent experience to the user, such as a financial service chatbot providing help to the branch customer.
    \item Tone and personality, such as a pessimist or an optimist.
    \item Proative or/and reactive nature depending on the purpose and viewpoint. proactive means actively taking the user towards a goal while reactive means letting the user lead the conversations by asking questions.
    \item Name of the chatbot which can be the first thing a user will notice in the chat box.
    \item Scenario describing brief narratives of anticipated use of a chatbot.
\end{itemize}

\textit{Responsible AI risk assessment pattern} needs to be applied to examine whether the purpose or viewpoint or scenario is unethical, whether the tone or name is offensive to a group of people. \textit{Ethical user story pattern} can be used to gather ethical requirements from stakeholders on the chatbot. An example user story can be  ``As an indigenous person, I want the chatbot to respond my questions in both English and indigenous language''. The ethical user stories can be used to trace the ethical requirements both backward to the stakeholders who developed the ethical user stories and forward into the design modules, code pieces, and test cases. \textit{Verifiable ethical requirement pattern} can be adopted to specify the ethical requirements in a verifiable form, e.g., ``all the responses must be provided with multiple language options including indigenous language option''.




\subsubsection{Conversation Design}
At the stage of conversation design, content writers, who are subject matter experts, perform the following tasks to provide the best user interaction based on the understanding of user behaviour. 
\begin{itemize}
    \item Further scenario development: The subject matter experts need to further define the scenarios that the chatbot is expected to fulfill.
    \item Question analysis: In this step, the subject matter experts need to define what are the key questions for the chatbot to answer within the scenario defined. These key questions are built as intents in the chatbot later. The subject matter experts also need to provide questions that chatbot developers can use to train the chatbot for each key question/intent. Ideally, the training data are expected to be collected from existing data sources such as call centre logs. However, these existing data are not easy to get. Thus, the subject matter experts need to script or create the questions. For example, super business users in the business domain are grouped together and asked to think how they would ask the question as if they ask the questions to their colleagues or think of how they are asked for such type of questions in their day to day job. This group of uses are then asked to write down the questions for each key questions.  
    \item Content design: After defining the key questions, the subject matter experts need to write responses to be displayed to user by the chatbot. 
    \item Conversation flow design: Once the content is completed, the subject matter experts need to design the dialogues that mimic human interaction. This step also needs to consider requirements on escalation to human agent.
    \item Interface design: Sometimes subject matter experts contribute to the design of user interface. 
\end{itemize}

\textit{Responsible AI risk assessment pattern} needs to be adopted to check if the defined scenarios describe any unethical narratives, whether the training data collected from existing data sources have any sensitive data or bias issue, whether the participated subject matter experts are diverse enough across gender, culture, race, age, expertise, etc.
\textit{Ethical user story pattern} can be used to collect the non-functional and ethical requirements for chatbot interface design, e.g., the chatbot interface design should consider the users who are colour blinded. 
\textit{Data requirement throughout the entire lifecycle pattern} is important to ensure that the data requirements are specified explicitly throughout the data lifecycle including the model training stage which may involve training data collected from existing data sources or third parties. 
\textit{Verifiable claim pattern} allows the developers to verify the ethical qualities of the training data which are associated with a verifiable claim on their ethical qualities.
\textit{Diverse team pattern} can be applied at the stage of conversation design. Building a diverse team of subject matter experts can effectively eliminate bias in responses and improve the design of dialogues.
\textit{Ethics training pattern} can be introduced to provide subject matter experts with knowledge and instructions on how to implement responsible AI in practice.
\textit{Human-centered interface design for explainable AI (XAI) pattern} is needed to provide helpful explanations to chatbot users, e.g., informing users that AI algorithms are used to generate responses.



\subsubsection{Implementation}
After completing the conversation design, these "raw materials" are handed over to a developer to configure into the chatbot platform, i.e., IBM Watson Assistant in this study. The developer needs to do the following configurations.
\begin{itemize}
\item Skill: A skill is created with the chatbot name.
\item Intents: Intents are configured for key questions provided by the subject matter experts. The intent is then trained with with utterances (sample questions) accordingly.
\item Entities: Entities are created to capture the context. For example, card type can be an entity class, while values can be credit card, debit card.
\item Dialogue: A dialogue is built to capture the responses in dialogue nodes. The conversation flow is configured by the flow control mechanism such as "jump to" and setting of Disambiguation or Digression. 
\item Advanced features: The developer could turn on the advanced features depending on the use case and business requirements, e.g., intent recommendations which use machine learning to uncover common topics in existing catalogues to quickly train the chatbot on the most frequent issues and questions; auto-learning that allows business to learn from customer choices to improve the journey.
\item Integration with downstream systems: The developer integrates the responses with downstream systems. 
\end{itemize}

\textit{Responsible AI risk assessment pattern} is needed to examine whether there is any misconfiguration by developers or machine learning algorithms, whether there is any vulnerability in data integration or storage, whether there is any privacy or bias issue with the sample questions for training.

\textit{Customised agile process pattern} can be used to adapt the agile development process by incorporating responsible AI  principles. Extension points could be artefacts, roles, ceremonies, practices, and culture. Ethical principles could be implemented through modifying the exiting artefacts (e.g., user stories) or adding new artefacts (e.g., regulatory requirements). There is also a need to promote responsible AI through existing roles (e.g., product) or new roles (e.g., ethicist). Similarly, modification of the existing ceremonies (e.g., sprint planning) or introduction of new ceremonies (e.g., ethics-oriented meetings) could be considered. Practices (e.g., user acceptance testing) and culture (e.g., hiring) are two effective ways to address ethical concerns in the agile development process.

\textit{Tight coupling of AI and non-AI development pattern} ensures the development of chatbot and other downstream systems are tightly coupled. The teams can share the same sprints and stand-up meetings, and use common co-versioning registry to manage the artifacts and track the progress. 

\textit{AI mode switcher pattern} is required for the configurations made by the machine learning algorithms. All the suggested configurations need to be reviewed and approved by developers before being released into production.

\textit{Data requirements throughout the entire lifecycle pattern} can be used to describe the requirements on data collection/integration and storage taking into account all the involved roles.

\textit{Co-versioning registry pattern} can be used to capture the relationships and dependencies of conversation design materials and implementation configurations.

\textit{Ethical construction with reuse pattern} allows the developers to reuse the previous configurations (e.g., intents) that have passed the responsible AI risk assessment.

\subsubsection{Testing}
This step is relatively straight forward. Conversations are tested with various testing methods such as unit testing, regression testing. 
\textit{Responsible AI risk assessment pattern} can be used to check if there is any ethical risks with the test cases. Particularly, \textit{ethical assessment for test cases} is required to assess all the test cases.
\textit{Ethical acceptance testing pattern} is needed to determine if the ethical requirements are met. Subject matter experts are involved in the testing process and perform the ethical acceptance testing. 

\subsubsection{Deployment}
In this step, the tested chatbot is deployed throughout the pipeline and released to production for end user to use. Oftentimes the final step to release to production needs to be approved by the chatbot owner or conversation owner.
\textit{Responsible AI risk assessment pattern} is needed to assess if there is any risk when integrating the chatbot with other systems and deploying the chatbot at scale.
\textit{Continuous deployment pattern} can be used to implement different deployment strategies. For example, phased deployment allows the chatbot to be only deployed for a subset group of users initially. The new version of chatbot rolls out incrementally and serves alongside the old version to reduce the ethical risk.
\textit{Standardized reporting pattern} is essential for governing AI systems. Organisations can setup standarized process and templates for informing the new release of chatbot to different stakeholders, such as financial service regulators and customers.

\subsubsection{Monitoring}
In this step, the performance of chatbot is measured by multiple metrics, such as Net Promoter Score (NPS) value, feedback from user, effectiveness and coverage of conversation. Most of the chatbot platforms provide analytic capability. Various metrics can be chosen and applied depending on business cases and needs. 
Also, this step ensures the chatbot answers the right questions, answers the questions in a right way, and answers enough questions. For the proactive chatbot, it means the chatbot needs to take the right actions.

\textit{Responsible AI risk assessment pattern} can be used to check if there are any missing or wrong metrics, whether there are any ethical issues in the questions from users and responses generated by chatbot, whether there is any bias in the results of user feedback.

\textit{Ethical black box pattern} can be used to continuously record the monitored data for improvement or auditing. Various approaches and methods can be applied to analyse the monitored data. Transaction review is one of the most popular methods for data analysis. When end users have conversation with the chatbot, the chat history can be captured and stored, e.g., by an immutable data ledger. 
Then all or a subset of the chat history can be review by the subject matter experts or auditors.

\textit{Independent oversight pattern} can be applied for auditing purposes 
Consistence issues need to be taken into consideration if multiple people perform the review as they may have different understanding. 

\textit{K-fold cross validation pattern} and \textit{blind testing pattern} can also be applied to evaluate the ethical performance of chatbot. These two patterns are new patterns we identified through the case study.

Once the analysis is completed, the developers configure and build the improvements required into the chatbot. The improvements can be as simple as retraining the chatbot or modifying the responses to provide more accurate and helpful responses in a responsible manner. It can also mean building new dialogue branches or responses that are required from the end users, or even integrating with new capability or other systems to provide more functionality for a new use case.

\section{CONCLUSION}

In this article, we discuss how we use a pattern-oriented responsible engineering approach to address the responsible AI challenges. We demonstrate the usefulness of the Responsible AI Pattern Catalogue in identifying and mitigating responsible AI risks through the chatbot development use case. We are currently examining the responsible AI risks of our internal AI projects across different domains and recommending pattern-driven migitations using the pattern catalogue. To automate the risk assessment process, we are developing a knowledge graph supported tool which can be used by different levels/types of stakeholders. The Responsible AI knowledge graph is constructed based on AI incidents database, our Responsible AI Pattern Catalogue and Question Bank.



\begin{thebibliography}{10}
\providecommand{\url}[1]{#1}
\csname url@samestyle\endcsname
\providecommand{\newblock}{\relax}
\providecommand{\bibinfo}[2]{#2}
\providecommand{\BIBentrySTDinterwordspacing}{\spaceskip=0pt\relax}
\providecommand{\BIBentryALTinterwordstretchfactor}{4}
\providecommand{\BIBentryALTinterwordspacing}{\spaceskip=\fontdimen2\font plus
\BIBentryALTinterwordstretchfactor\fontdimen3\font minus
  \fontdimen4\font\relax}
\providecommand{\BIBforeignlanguage}[2]{{%
\expandafter\ifx\csname l@#1\endcsname\relax
\typeout{** WARNING: IEEEtran.bst: No hyphenation pattern has been}%
\typeout{** loaded for the language `#1'. Using the pattern for}%
\typeout{** the default language instead.}%
\else
\language=\csname l@#1\endcsname
\fi
#2}}
\providecommand{\BIBdecl}{\relax}
\BIBdecl

\bibitem{jobin2019global}
A.~Jobin, M.~Ienca, and E.~Vayena, ``The global landscape of ai ethics
  guidelines,'' \emph{Nature Machine Intelligence}, vol.~1, no.~9, pp.
  389--399, 2019.

\bibitem{liao2020questioning}
Q.~V. Liao, D.~Gruen, and S.~Miller, ``Questioning the ai: informing design
  practices for explainable ai user experiences,'' in \emph{Proceedings of the
  2020 CHI Conference on Human Factors in Computing Systems}, 2020, pp. 1--15.

\bibitem{han2022checklist}
S.-H. Han and H.-J. Choi, ``Checklist for validating trustworthy ai,'' in
  \emph{2022 IEEE International Conference on Big Data and Smart Computing
  (BigComp)}.\hskip 1em plus 0.5em minus 0.4em\relax IEEE, 2022, pp. 391--394.

\bibitem{raji2020closing}
I.~D. Raji, A.~Smart, R.~N. White, M.~Mitchell, T.~Gebru, B.~Hutchinson,
  J.~Smith-Loud, D.~Theron, and P.~Barnes, ``Closing the ai accountability gap:
  Defining an end-to-end framework for internal algorithmic auditing,'' in
  \emph{Proceedings of the 2020 conference on fairness, accountability, and
  transparency}, 2020, pp. 33--44.

\bibitem{hutchinson2021towards}
B.~Hutchinson, A.~Smart, A.~Hanna, E.~Denton, C.~Greer, O.~Kjartansson,
  P.~Barnes, and M.~Mitchell, ``Towards accountability for machine learning
  datasets: Practices from software engineering and infrastructure,'' in
  \emph{Proceedings of the 2021 ACM Conference on Fairness, Accountability, and
  Transparency}, 2021, pp. 560--575.

\bibitem{lu2022towards}
Q.~Lu, L.~Zhu, X.~Xu, J.~Whittle, and Z.~Xing, ``Towards a roadmap on software
  engineering for responsible ai,'' in \emph{Proceedings of the 1st
  International Conference on AI Engineering: Software Engineering for AI},
  2022, pp. 101--112.

\bibitem{lu2022responsible}
Q.~Lu, L.~Zhu, X.~Xu, J.~Whittle, D.~Zowghi, and A.~Jacquet, ``Responsible ai
  pattern catalogue: a multivocal literature review,'' \emph{arXiv preprint
  arXiv:2209.04963}, 2022.

\bibitem{lu2023responsible}
Q.~Lu, L.~Zhu, X.~Xu, and J.~Whittle, ``Responsible-ai-by-design: A pattern
  collection for designing responsible ai systems,'' \emph{IEEE Software},
  2023.

\bibitem{weizenbaum1976computer}
J.~Weizenbaum, ``Computer power and human reason: From judgment to
  calculation.'' 1976.

\bibitem{shin2022cross}
D.~Shin, S.~Al-Imamy, and Y.~Hwang, ``Cross-cultural differences in information
  processing of chatbot journalism: chatbot news service as a cultural
  artifact,'' \emph{Cross Cultural \& Strategic Management}, vol.~29, no.~3,
  pp. 618--638, 2022.

\bibitem{shin2022perception}
D.~Shin, ``The perception of humanness in conversational journalism: An
  algorithmic information-processing perspective,'' \emph{New Media \&
  Society}, vol.~24, no.~12, pp. 2680--2704, 2022.

\bibitem{shin2022effects}
D.~Shin, V.~Chotiyaputta, and B.~Zaid, ``The effects of cultural dimensions on
  algorithmic news: How do cultural value orientations affect how people
  perceive algorithms?'' \emph{Computers in Human Behavior}, vol. 126, p.
  107007, 2022.

\bibitem{shin2022algorithm}
D.~Shin, K.~F. Kee, and E.~Y. Shin, ``Algorithm awareness: Why user awareness
  is critical for personal privacy in the adoption of algorithmic platforms?''
  \emph{International Journal of Information Management}, vol.~65, p. 102494,
  2022.

\bibitem{shin2022understanding}
D.~Shin, J.~S. Lim, N.~Ahmad, and M.~Ibahrine, ``Understanding user sensemaking
  in fairness and transparency in algorithms: algorithmic sensemaking in
  over-the-top platform,'' \emph{AI \& SOCIETY}, pp. 1--14, 2022.

\bibitem{shin2022platforms}
D.~Shin, B.~Zaid, F.~Biocca, and A.~Rasul, ``In platforms we trust? unlocking
  the black-box of news algorithms through interpretable ai,'' \emph{Journal of
  Broadcasting \& Electronic Media}, vol.~66, no.~2, pp. 235--256, 2022.

\bibitem{shin2022seeing}
D.~Shin, A.~Rasul, and A.~Fotiadis, ``Why am i seeing this? deconstructing
  algorithm literacy through the lens of users,'' \emph{Internet Research},
  vol.~32, no.~4, pp. 1214--1234, 2022.

\bibitem{lu2022responsible1}
Q.~Lu, L.~Zhu, X.~Xu, and J.~Whittle, ``Responsible-ai-by-design: a pattern
  collection for designing responsible ai systems,'' \emph{arXiv preprint
  arXiv:2203.00905}, 2022.

\bibitem{zhu2022ai}
L.~Zhu, X.~Xu, Q.~Lu, G.~Governatori, and J.~Whittle, ``Ai and
  ethics—operationalizing responsible ai,'' in \emph{Humanity Driven
  AI}.\hskip 1em plus 0.5em minus 0.4em\relax Springer, 2022, pp. 15--33.

\end{thebibliography}


\end{document}